\documentclass[11pt,twoside]{article}

\usepackage[ansinew]{inputenc} 
\usepackage{amsmath,amsfonts,amssymb,amsthm}
\usepackage{pstricks,pst-node,pst-coil,pst-plot,pstricks-add}
\usepackage{geometry,epsfig}
\usepackage{bbm}
\usepackage{cite}
\usepackage{graphicx}

\usepackage{bbm}
\usepackage{cite}
\usepackage{paralist}
\usepackage{cleveref}
\usepackage{enumitem}
\usepackage{emptypage}

\DeclareGraphicsExtensions{.pdf}

\usepackage{mathrsfs} 

\usepackage[T1]{fontenc}

\newcommand{\field}[1]{\mathbb{#1}}
\newcommand{\N}{\field{N}}
\newcommand{\R}{\field{R}}

\newcommand{\HH}{\mathscr H}
\newcommand{\KK}{\mathscr K}

\newcommand{\eps}{\varepsilon}
\newcommand{\ph}{\varphi}

\newcommand{\restricted}{|\grave{}\,}

\newcommand{\sprod}[2]{\langle #1,#2 \rangle}        

\newcommand{\sgn}{\operatorname{sgn}}
\newcommand{\supp}{\operatorname{supp}}

\newcommand{\diam}{\operatorname{diam}}
\newcommand{\dist}{\operatorname{dist}}

\newcommand{\Hepsf}{H_{\eps}}

\newcommand{\HHf}{\mathscr{H}_{\textup{f}}}

\newcommand{\XXf}{\mathfrak X_{\textup{f}}}

\newcommand{\lambdamax}{\lambda_{\textup{max}}}
\renewcommand{\d}[1]{\textup{d}#1}

\newtheorem{thm}{Theorem}[section]
\newtheorem{prop}[thm]{Proposition}

\newtheorem{lemma}[thm]{Lemma}

\title{\textbf{On the Weakness of Short-Range Interactions\\ in Fermi Gases}}

\author{M.~Griesemer\footnote{marcel.griesemer@mathematik.uni-stuttgart.de}\,\ and M.~Hofacker\footnote{michael.hofacker@kit.edu}\\  
\small Fachbereich Mathematik, Universit\"at Stuttgart, D-70569 Stuttgart, Germany}  
\date{}

\begin{document}
\maketitle

\begin{abstract}
Ultracold quantum gases of equal-spin fermions with short-range interactions are often considered free even in the presence of strongly binding spin-up-spin-down pairs. We describe a large class of many-particle Schr\"odinger operators with short-range pair interactions, where this approximation can be justified rigorously.
\end{abstract}


\section{Introduction}
Short-range interactions among equal spin fermions in ultracold quantum gases are often neglected, while at the same time the interaction between particles of opposite spin is modeled by zero-range (i.e.~contact) interactions \cite{Chevy2006,RevModPhys2008,Parish2011}. This can be justified by the fact that zero-range interactions among spinless (or equal spin) fermions are prohibited by the Pauli principle (see Theorem~\ref{thm:Pauli}, below), and by the recent approximation results for zero-range interactions in terms of short-range potentials  \cite{Basti2018,GHL2019,GH2021}. 
In the present paper we give a more direct analysis of the weakness of short-range interactions among spinless fermions in terms of estimates for the resolvent difference of free and interacting Hamiltonians. Our main results hold for all space dimensions $d\leq 3$.

We consider fermionic $N$-particle systems in the Hilbert space 
$$
       \HH_f = \wedge ^N L^2(\R^d)
$$
described by Schr\"odinger operators
\begin{equation}\label{intro1}
       H_\eps = -\Delta - \lambda_\eps \sum_{i<j}V_\eps(x_i-x_j),
\end{equation}
where $V_\eps(r) = \eps^{-2}V(r/\eps)$, $V(-r) = V(r)$, $\lambda_\eps>0$ and $d\in\{1,2,3\}$. We are primarily interested in the case, where the interaction strength of two distinct particles, in their center-of-mass frame described by
\begin{equation}\label{intro2}
        -2\Delta  - \lambda_\eps V_\eps,\quad \text{in}\ L^2(\R^{d}),
\end{equation}
is independent of $\eps$ in the sense that \eqref{intro2} has a ground state energy $E_{\eps}$ that is fixed or convergent  $E_{\eps}\to E$ with limit $E<0$.  It is well-known in the spectral theory of Schr\"odinger operators what this means for $\lambda_\eps$ \cite{Simon76,KlausSimon1980}. In fact, $\eps\mapsto \lambda_\eps$ and $V$ can be chosen, depending on $d$, in such a way that the resolvent of \eqref{intro2} -- and hence the spectrum of \eqref{intro2} -- has a limit as $\eps\to 0$. Since $E<0$, the limit of \eqref{intro2}  describes a non-trivial point interaction at the origin \cite{SolvableModels}.

Our main result, \Cref{thm:main}, can be described in simplified form as follows. Assuming $V\in L^1\cap L^2(\R^d)$ with $V\geq 0$,  $C_V = \sup_{r\in \R^d}V(r)|r|^2<\infty$, some further decay of $V$ in the case $d=1$, and 
\begin{equation}\label{intro3}
    \limsup_{\eps\to 0} \lambda_\eps <\frac{d^2}{C_VN},
\end{equation}
we show that 
\begin{equation}\label{intro4}
   H_\eps \to -\Delta \qquad (\eps\to 0)
\end{equation}
in norm resolvent sense. The rate of convergence depends on the size of $\lambda_\eps$ and - to some extent - on the decay of $V(x)$ as $|x|\to \infty$. Given sufficient decay of $V$, the regularity of $H^2(\R^d)$-functions, and hence the dimension $d$, begins to play a role. If we choose $\lambda_\eps$ as described above, where $E_\eps\to E<0$, then condition \eqref{intro3} is satisfied for all $N$ if $d\leq 2$, and for some $N\geq 3$ if $d=3$. Surprisingly, this particular choice of $\lambda_\eps$ conspires with the regularity of Sobolev functions in such a way that 
\begin{equation}\label{intro5}
      \|(H_\eps +z)^{-1} - (-\Delta+z)^{-1} \| = O(\eps^2)\qquad (\eps\to 0)
\end{equation}
with the bound $O(\eps^2)$ independent of the space dimension $d$. Another choice for the coupling constant, consistent with \eqref{intro3} as well, is the one where $\lambda_\eps$ is a positive constant smaller than $d^2/C_VN$. Then operator \eqref{intro2}, upon a rescaling, is proportional to $\eps^{-2}$ and hence the (negative) binding energy $E_\eps$ diverges. In this case, the limit $\eps\to 0$ amounts to a combined short-range and strong interaction limit, which is interesting and relevant physically \cite{Parish2011}.

In summary we can say that fully spin polarized Fermi gases in $d\leq 2$ with short-range interactions - the spin-up-spin-down interaction strength being fixed - are asymptotically free in the limit of zero-range interaction. This is true in dimension $d=3$ as well for suitable $V$ and small $N\geq 3$, depending on $V$. The result remains correct even in a suitable combined limit of short-range and strong interaction.

We conclude with some remarks on the literature:  For single particles the approximation of point-like disturbances by short-range potentials is discussed at length and in rich detail in \cite{SolvableModels}, see also \cite{BHS2013}. 
For systems of $N\geq 3$ particles in $d\leq 2$ dimensions, it was recently shown in a series of papers, that contact interactions (of TMS-type) can be approximated by rescaled two-body potentials in the norm resolvent sense of $N$-particle Hamiltonians \cite{Basti2018,GHL2019,Quastel2021,GH2021,Hofacker-Diss}. From these results the mere convergence \eqref{intro4} can be derived by reduction of the Hilbert space to antisymmetric wave functions.  This works for the very special choice of $\lambda_\eps$ needed for the approximation of contact interactions, and for $d\leq 2$, only. For $d=3$, systems of $N\geq 3$ distinct (or bosonic) particles with two-body short-range interactions are prone to suffer collapse, a phenomenon known as Thomas effect. See \Cref{thomas}, below. In order to avoid this effect, suitable many-body forces are required
\cite{Basti2021, FerrTeta22a, FerrTeta22b}.

This work is organized as follows. In Section \ref{sec:two} we present an explicit estimate of the norm of $(H_{\eps=1} +z)^{-1} - (-\Delta+z)^{-1}$ in terms of the pair potential $V$. This estimate is then used in Section \ref{sec:three} to prove \eqref{intro4} and \eqref{intro5}. The proofs benefit from the methods and tools developed in \cite{GHL2019,GH2021}. Section \ref{sec:four} gives examples illustrating our results, in particular for $d=3$ and $N=3$. Finally, in Section \ref{sec:appendix}, we prove the impossibility of fermionic contact interaction in $d\geq 2$. This improves, for fermions, a well known result about the impossibility of  contact interactions in dimensions $d\geq 4$ \cite{Sven1981}.


\section{The resolvent difference}
\label{sec:two}

Let $H_0=-\Delta$ in $\HH_f$ and let $R_0(z) = (H_0+z)^{-1}$. We consider $N$-body Hamiltonians  $H= H_0 - W$ in $\HH_f$ with
 $$W = \sum_{i<j} V_{ij},$$ 
where $V_{ij}$ denotes multiplication with $V(x_i-x_j)$ and $V\in L^1\cap L^2(\R^d)$ is real-valued and even. There is no scaling parameter and no coupling constant. We assume $d\leq 3$ and hence $H$ is self-adjoint on $D(H)=D(H_0)$ \cite{RS2}. The result of this section, \Cref{res-diff}, is based on a suitable factorization $W = A^{*}B$ and an iterated (second) resolvent identity related to the Konno-Kuroda formula \cite{KoKu1966}. We start by constructing the factorization.

Let
\begin{align}
    \XXf:= L^2_{\textup{odd}}(\R^d, \d{r}) \otimes  L^2(\R^d, \d{R}) \otimes \bigwedge_{i=3}^{N} L^2(\R^d,\textup{d}x_i), \label{DefXfermi}
\end{align}
where $L^2_{\textup{odd}}$ denotes the subspace of odd functions from $L^2$. The integration  variables $r$ and $R$ in \eqref{DefXfermi} correspond to the relative and center of mass coordinates of the fermion positions $x_1$ and $x_2$. This change of coordinates is implemented isometrically by the operator $\KK: \HHf \rightarrow \XXf$ given by
\begin{equation} \label{DefK}
         \big(\KK\psi \big)\left(r,R,x_3,...,x_N\right):= \psi\left(R-\frac{r}{2},R+\frac{r}{2}, x_3,..., x_{N}\right).
\end{equation}

From $\sprod{\ph}{V_{ij}\psi} = \sprod{\ph}{V_{12}\psi}$ and from $\sprod{\ph}{V_{12}\psi}=\sprod{\KK\ph}{\KK V_{12}\psi} = \sprod{\KK\ph}{(V\otimes 1)\KK\psi}$ it follows that 
\begin{equation} \label{Potdecomp}
       W = {N\choose 2}\KK^{*}(V\otimes 1)\KK.
\end{equation} 
We therefore write  $V = vu$ with 
\begin{align*}
      v(r) &:= |V(r)|^{1/2}, \\
      u(r) &:= J |V(r)|^{1/2},\qquad J:=\sgn(V),
\end{align*}
and we set 
\begin{align}
    A &:= {N\choose 2}^{1/2} (v \otimes 1)  \KK, \label{DefA} \\
    B &:= {N\choose 2}^{1/2} (u \otimes 1) \KK = JA. \label{DefB}
\end{align}
Recall that $V \in L^1(\R^d)$ and hence $u, v \in L^2(\R^d)$. The domain $D(A)$ of $A: D(A) \subset \HHf \rightarrow \XXf$ is determined by the domain of the multiplication operator $v \otimes 1$, so it follows that $A$ and $B$ are densely defined and closed on $D(A) \supset D(H_0)$. 
Hence, $H$ can be rewritten in the form
\begin{equation}\label{Hepsnew}
     H= H_0 - A^{*} B. 
\end{equation}

By an iteration of the resolvent identity, we find
\begin{equation}\label{res-id}
    (H + z)^{-1}  = R_0(z) + R_0(z)WR_0(z) + R_0(z)W (H + z)^{-1} WR_0(z).
\end{equation}
Upon setting  $W=A^{*}B$, Identity~\eqref{res-id} can be written in the form
\begin{align}
         (H + z)^{-1}  &= R_0(z) + (AR_0(\overline{z}))^{*} S(z) BR_0(z),\label{KKformula}\\
         S(z) &= 1 + B(H+z)^{-1}A^{*}.\label{Lambda-1}
\end{align}
The following proposition is our tool for proving norm resolvent convergence in the next section. 
\begin{prop}\label{res-diff}
Suppose there exists $\delta>0$ such that $H_0 - (1+\delta)W\geq 0$ and suppose that $V\geq 0$, $V\leq 0$, or that
 $\sum_{i<j} |V_{ij}| \leq CH_0$ with some $C>0$. Then $H\geq 0$ and, for all $z>0$,
$$
       \|(H+z)^{-1} - (H_0+z)^{-1}\| \leq C_\delta {N \choose 2} \|v(-\Delta+z)^{-1}\|^2_{\rm odd},
$$
where  $v = |V|^{1/2}$, and $C_\delta$ is a function of $C$ and $\delta$. Here $\|\cdot\|_{\rm odd}$ denotes the operator norm in $L^2_{\textup{odd}}(\R^d, \d{r})$.
\end{prop} 

\begin{proof}
Suppose, temporarily, that $H\geq 0$ and let $z>0$. Then, from \eqref{KKformula}, the definition of $A$, and from $\|(-\Delta_r\otimes 1+z)\KK(H_0+z)^{-1}\|\leq 1$, it follows that
$$
    \|(H+z)^{-1} - R_0(z)\| \leq {N\choose 2} \|v(-\Delta+z)^{-1}\|^2_{\rm odd} \|S(z)\|. 
$$
It remains to prove $H\geq 0$ and $\|S(z)\|\leq C_\delta$ under the various assumptions on $V$.

In the case $V\leq 0$ we have $J=-1$, $B=-A$, and hence, $H=H_0+ A^{*}A \geq A^{*}A \geq 0$. By \Cref{lm:basic} below, it follows that 
$\|S(z)\| \leq 2$.

In the case $V\geq 0$ we have $J=1$, $B=A$, and hence the assumption $H_0 - (1+\delta)W\geq 0$ implies that 
$$
       H\geq \delta W = \delta A^{*}A.
$$
By \Cref{lm:basic} it follows that $A(H+\mu)^{-1}A^{*} \leq 1/\delta$, and hence $\|S(z)\|\leq 1 +\delta^{-1}$, by \eqref{Lambda-1}.

It remains to consider the case where $V$ changes sign. The assumption $H_0 - (1+\delta)W\geq 0$ implies that 
$$
    H = \frac{\delta}{1+\delta}H_0 + \frac{1}{1+\delta}(H_0 - (1+\delta)W) \geq  \frac{\delta}{1+\delta}H_0.
$$
Combining this with the assumption on $|V|$, that is, with $CH_0 \geq \sum_{i<j} |V_{ij}| = A^{*}A$, we find that
$$
     H \geq  \frac{\delta}{1+\delta} \frac{1}{C} A^{*}A.
$$
This, by \Cref{lm:basic}, implies
$A(H+\mu)^{-1}A^{*} \leq C (1+\delta)/\delta$ and the desired bound on $\|S(z)\|$ follows from \eqref{Lambda-1}.
\end{proof}

\begin{lemma}\label{lm:basic}
Let $H, A$ be any two closed operators in a Hilbert space, with $H^{*}=H\geq 0$ and $D(A)\supset D(H)$. If
$H\geq \lambda A^{*}A$ for some $\lambda>0$, then  $D(A)\supset D(H^{1/2})$ and for all $\mu>0$, 
$$
     A(H+\mu)^{-1}A^{*} \leq \frac{1}{\lambda}.
$$ 
\end{lemma}

\begin{proof}
The assumption means that $\|(H+\mu)^{1/2}\psi\|^2 \geq \lambda\|A\psi\|^2$ for $\mu>0$ and all $\psi\in D(H)$.
By an approximation argument this inequality extends to all $\psi\in D(H^{1/2})$, and $D(A)\supset D(H^{1/2})$ follows from the closedness of $A$. Upon rewriting the inequality in terms of $\ph = (H+\mu)^{1/2}\psi\in \HH_f$, the assertion follows. 
\end{proof}


\section{The resolvent convergence}
\label{sec:three}

We now apply the results from the previous section to Schr\"odinger operators with rescaled two-body potentials, that is 
\begin{equation}\label{schrodinger}
       H_\eps = -\Delta - \lambda_\eps \sum_{i<j}V_{\eps,ij},
\end{equation}
where $\lambda_\eps>0$ and $V_\eps(r) = \eps^{-2}V(r/\eps)$. The Schr\"odinger operator  $\eps^2 H_\eps$  is unitarily equivalent to $S_{\lambda} := -\Delta -\lambda\sum_{i<j}V_{ij}$ with $\lambda = \lambda_\eps$. We therefore define,
\begin{equation}\label{lambda-max}
    \lambdamax:=\sup\{\lambda\geq 0 \mid S_\lambda\geq 0\}.
\end{equation}
Note that $S_\lambda\geq 0$ for all $\lambda\leq  \lambdamax$. This follows from the fact that $E_\lambda:=\inf\sigma(S_\lambda)$ as a function of $\lambda$ is concave (hence continuous) and $E_0=0$. 

For \Cref{thm:main} to be non-void, we need that $ \lambdamax>0$. This can be achieved,  e.g., by assuming that 
$$
     C_V:=\sup_{x\in \R^d} V(x)|x|^2 <\infty.
$$
Then $\lambdamax \geq d^2/(C_VN)>0$, by the Hardy inequality for fermionic wave functions $\psi\in \HH_f$ \cite{HOLT2007},
\begin{equation}\label{hardy}
       \sum_{i<j}\int\frac{|\psi(x_1,\ldots,x_N)|^2}{|x_i-x_j|^2}\,dx \leq \frac{N}{d^2}\|\nabla\psi\|^2.
\end{equation}

Statement as well as proof of \Cref{thm:main} depend on the regularity of $H^2$-Sobolev functions. Explicitly we use the embedding $H^2(\R^d) \hookrightarrow C^{0,s}(\R^d)$, valid for $s\in I_d$, where  
$I_1=[0,1], I_2=[0,1)$ and $I_3=[0,1/2]$, and we use \Cref{log-holder}, below, which improves the embedding in the case $d=2$. 
Here $ C^{0,s}(\R^d)$ denotes the space of continuous functions that are uniformly H\"older continuous of exponent $s$.

\begin{thm}\label{thm:main}
Suppose that $V\geq 0$, $V\leq 0$, or that $\sup_{x\in \R^d} |V(x)||x|^2 <\infty$.
If $\lambda_0:= \limsup_{\eps\to 0} \lambda_{\eps} <\lambdamax $, then $H_\eps\geq 0$ for $\eps$ small enough, and for all $z>0$,
\begin{equation}\label{free-limit}
   \|(H_\eps+z)^{-1} - (H_0+z)^{-1}\| = o(\lambda_\eps \eps^{d-2}) \qquad (\eps\to 0).
\end{equation}
Moreover, the following is true:
\begin{itemize}
\item[(a)] If $\int |V(r)| |r|^{2s}\, dr<\infty$ for some $s\in I_d$, then
   $$ \|(H_\eps+z)^{-1} - (H_0+z)^{-1}\| = O(\lambda_\eps \eps^{d-2+2s})\qquad (\eps\to 0).$$
\item[(b)] If $d=2$ and $\int |V(r)| |r|^{2}|\log|r||\, dr<\infty$, then
   $$ \|(H_\eps+z)^{-1} - (H_0+z)^{-1}\| = O(\lambda_\eps \eps^{d}|\log\eps|)\qquad (\eps\to 0).$$
\end{itemize}
\end{thm}


In the situation described in the introduction, where $\inf\sigma(-2\Delta - \lambda_\eps V_\eps)$ has a limit $E<0$, the bounds (a) and (b) reveal a surprising interplay between $\lambda_\eps$ and the (optimal) regularity of $H^2(\R^d)$-functions: if, depending on $d$, we choose $\lambda_\eps = O(\eps)$, $\lambda_\eps = O(1/|\log\eps|)$, and $\lambda_\eps = O(1)$ for $d=1$, $d=2$, and $d=3$, respectively, then, for all $d\in \{1,2,3\}$,
$$
         \|(H_\eps+z)^{-1} - (H_0+z)^{-1}\| = O(\eps^2),\qquad  (\eps\to 0)
$$
provided $V$ decays fast enough, e.g., as in the hypothesis of (b), and $N$ is small if $d=3$.

\noindent
\emph{Remarks:} 
\begin{enumerate}
\item Part (a) of the theorem shows that, for a large class of potentials,
\begin{equation}\label{ham-rc}
     H_\eps \to H_0\qquad (\eps\to 0)
\end{equation}
in the norm resolvent sense, provided that $\lambda_\eps \eps^{d-2+2s} \to 0$ as $\eps\to 0$. This is not true for the Hamiltonians $\tilde{H}_\eps$ defined by \eqref{schrodinger} on the enlarged Hilbert space $L^2(\R^{Nd})$, as shown in the next section.

\item For the convergence \eqref{ham-rc} in norm resolvent sense to hold, it is necessary that $\inf\sigma(H_\eps) \to \inf\sigma(H_0) = 0$. Therefore the assumption $ \limsup_{\eps\to 0} \lambda_{\eps} <\lambdamax$ in \Cref{thm:main} cannot be relaxed significantly. Strong resolvent convergence, by contrast, has much weaker spectral implications and hence - given some decay of $V$ - much less is needed of $\lambda_\eps$, see \Cref{prop:strong} below.

\item  A weaker result, similar to \Cref{thm:main}, could be derived from \cite{GHL2019, GH2021}. Indeed, for suitable $\lambda_\eps$ we know from \cite{GHL2019, GH2021} that $H_\eps\to H$ in norm resolvent sense, where $H= -\Delta$ on $\HHf$. Information on the rate of convergence can also be found in these papers.
\end{enumerate}

\begin{proof}
We are going to apply \Cref{res-diff} to the Hamiltonian \eqref{schrodinger}, and we assume that $V$ changes sign, the other cases being easier. Due to the unitary equivalence of $\eps^2 H_\eps$ and $S_{\lambda_\eps}$, 
the hypotheses of \Cref{res-diff} are equivalent to
\begin{align}
      H_0 - (1+\delta)\lambda_\eps\sum_{i<j}V_{ij} &\geq 0 \label{hyp1}\\
      CH_0 - \lambda_\eps\sum_{i<j}|V_{ij}| &\geq 0 \label{hyp2}
\end{align}
for some $\delta,C>0$. Both \eqref{hyp1} and \eqref{hyp2} are true for $\eps$ small enough. This follows from $\lambda_0=\limsup_{\eps\to 0}\lambda_\eps <\lambdamax$, from $\sup_{x\in \R^d} |V(x)||x|^2 <\infty$, and from the Hardy inequality for fermions \eqref{hardy}. Hence, by \Cref{res-diff}, for $\eps>0$ small enough,
\begin{equation}\label{step1}
       \|(H_\eps+z)^{-1} - (H_0+z)^{-1}\| \leq  C_\delta {N \choose 2}  \lambda_\eps\eps^{d-2} \|v_\eps(-\Delta+z)^{-1}\|^2_{\rm odd},
\end{equation}
where $v_\eps(x) := \eps^{-d/2}|V(x/\eps)|^{1/2}$.

By the Sobolev embedding $H^2(\R^d) \hookrightarrow C^{0,s}(\R^d)$, valid for $s\in I_d$, the elements $\psi \in H^2(\R^d) \cap L^2_{\textup{odd}}(\R^d)$ are H\"older continuous (of exponent $s$) odd functions. It follows that $\psi(0)=0$ and that 
\begin{equation}\label{use-holder}
     |\psi(x)| = |\psi(x)-\psi(0)| \leq C_s\|\psi\|_{H^2} |x|^s.
\end{equation}
Therefore, for all $\psi \in H^2(\R^d) \cap L^2_{\textup{odd}}(\R^d)$, 
\begin{align}
    \|v_\eps\psi\|^2 &\leq C_s^2 \|\psi\|_{H^2}^2 \int |v_\eps(x)|^2 |x|^{2s}\, dx\nonumber \\
             &=  \eps^{2s} C_s^2 \|\psi\|_{H^2}^2 \int |V(x)| |x|^{2s}\, dx. \label{vpsi-bound}
\end{align}
This is true for all $s\in I_d$ and, combined with \eqref{step1}, it proves statement (a) of the theorem. To prove (b), we use 
\Cref{log-holder} in \eqref{use-holder} (rather than $H^2 \hookrightarrow C^{0,s}$) and then \eqref{vpsi-bound} becomes $C\eps^2 |\log\eps|\int |V(x)| |x|^2(1+|\log|x||)\, dx$, where the integral is finite by the assumptions on $V$. 

It remains to prove \eqref{free-limit}. Eq.~\eqref{vpsi-bound} with $s=0$ implies that $\|v_\eps(-\Delta+z)^{-1}\|_{\rm odd} = O(1)$, 
which can be improved as follows: let $\chi_k$ denote the characteristic function of the ball $|x|\leq k$ in $\R^d$ and let 
$(v\chi_k)_\eps = v_\eps\chi_{\eps k}$. Then, by \eqref{vpsi-bound}, 
\begin{equation}\label{vk-bound}
     \|(v\chi_k)_\eps(-\Delta+z)^{-1}\|^2_{\rm odd} = O(\eps^{2s}) = o(1)\qquad (\eps\to 0)
\end{equation}
for any $s>0$ in $I_d$. On the other hand, 
\begin{equation}\label{v-vk-bound}
     \|(v - v\chi_k)_\eps(-\Delta+z)^{-1}\|_{HS} \leq C\|v - v\chi_k\| = o(1)\qquad (k\to \infty)
\end{equation}
uniformly in $\eps>0$, where $HS$ refers to Hilbert-Schmidt norm. The combination of \eqref{step1}, \eqref{vk-bound} and \eqref{v-vk-bound} proves \eqref{free-limit} and concludes the proof of the theorem. 
\end{proof}


In the proof of \Cref{thm:main} we have used the following lemma, which can probably be found in the literature, but we are not aware of suitable reference.

\begin{lemma}\label{log-holder}
For all $u\in H^2(\R^2)$ and all $x,y\in\R^2$, $y\neq 0$, we have
$$
        |u(x+y) - u(x)| \leq \frac{1}{2\sqrt{\pi}}|y| \big(2+|\log|y||\big)^{1/2} \left(\|\Delta u\|^2 + \|\nabla u\|^2\right)^{1/2}.
$$
\end{lemma}

\noindent
\emph{Remark:} By our method of proof, this inequality can be generalized to derivatives $\partial^{\alpha} u$, $|\alpha|\leq k$, of functions  $u\in H^s(\R^n)$ with $s-(n/2) = k+1$, $s\in\R$ and $k\in \N_0$. 

\begin{proof} We first note that $u \in H^2(\R^2)$ implies that $\widehat{u} \in L^1(\R^2)$, and hence, for all $x \in \R^2$,
\begin{align}
u(x) = \dfrac{1}{2\pi} \int_{\R^2} \widehat{u}(p) \exp(ipx)  \, \d{p}.
\nonumber 
\end{align}
Therefore, by Cauchy-Schwarz,
\begin{align}
2\pi \dfrac{|u(x+y) - u(x) | }{|y|} &\leq 
\int_{\R^2} \dfrac{|\exp(ipy)- 1|}{|y|} |\widehat{u}(p)|  \, \d{p}\nonumber\\ &
\leq I(y)^{1/2}  \left( \| \Delta u \|^2 + \| \nabla u \|^2  \right)^{1/2},
\label{Hoelderabs2}
\end{align}
where
\begin{align}
I(y):= \int_{\R^2} \dfrac{|\exp(ipy)- 1|^2}{|y|^2} \dfrac{1}{|p|^4 + |p|^2} \, \d{p}. \nonumber
\end{align}
To estimate the integral $I(y)$, we may assume that $y=(|y|,0)$. Then, upon the substitution $q=p|y|$, we find that, for any $Q>0$,
\begin{equation}\label{I(y)est}
I(y) = \int\limits_{\R^2}  \dfrac{|\exp(iq_1)- 1|^2}{|q|^4 + |y|^2|q|^2} \, \d{q} \leq  \int\limits_{|q| \leq Q} \dfrac{q_1^2}{|q|^4 + |y|^2|q|^2} \, \d{q} +  \int\limits_{|q| > Q} \dfrac{4}{|q|^4} \, \d{q},  
\end{equation}
where $|\exp(iq_1)- 1| \leq |q_1|$ and $|\exp(iq_1)- 1| \leq 2$ was used, respectively. Both integrals on the right of  \eqref{I(y)est} can be computed explicitly. For the first one we obtain
\begin{align}
\int\limits_{|q| \leq Q} \dfrac{q_1^2}{|q|^4 + |y|^2|q|^2} \, \d{q}
&= \dfrac{1}{2} \int\limits_{|q| \leq Q} \dfrac{1}{|q|^2 + |y|^2} \, \d{q}  \nonumber  \\
&= \dfrac{\pi}{2} \log\left(1+ \dfrac{Q^2}{|y|^2} \right) \leq \pi \left( 
|\log|y|| + |\log Q| + \frac{1}{2} \log(2) \right),
\label{FirstEst}
\end{align}
where the inequality follows from  $\log(1+t) \leq |\log t| + \log 2$, valid for all $t>0$.
The second integral on the right side of \eqref{I(y)est} equals $4 \pi/Q^2$. Choosing $Q=2\sqrt{2}$, we find from \eqref{I(y)est} and \eqref{FirstEst} that
$$
   I(y) \leq \pi (|\log|y|| + c)
$$
with $c=\frac{1}{2} + \log 4 < 2$. This concludes the proof.
\end{proof}


We conclude this section with the proposition announced in Remark 2, above. It is a consequence of \Cref{thm:Pauli} concerning the essential self-adjointness of the Laplacian. In all the following $\Omega = \R^{Nd} \setminus \Gamma$,  where 
\begin{equation}\label{def-Gamma}
     \Gamma := \bigcup_{i<j}\big\{x=(x_1,\ldots,x_N)\in \R^{Nd} \mid x_i=x_j\big\}.
\end{equation}
Furthermore, $C_0^{\infty}(\Omega):=\{\psi\in C_0^{\infty}(\R^{Nd})\mid \supp\psi\subset \Omega\}$.

\begin{prop}\label{prop:strong}
Let $d\geq 2$ and suppose that $V\in L^2(\R^d)$. Suppose there exists $s\geq 0$ such that $\int V(r)^2 |r|^{2s}\,\d{r} < \infty$ and $\limsup_{\eps \to 0} \lambda_{\eps}\,\eps^{s+d/2 - 2} < \infty$. Then, $H_\eps \to H_0$ in the strong resolvent sense as $\eps\to 0$.
\end{prop}

\begin{proof}
In view of  $\|(\Hepsf+i)^{-1}\|=1=\|R_0(i)\|$ it suffices to prove that $(H_\eps+i)^{-1}\psi \to R_0(i)\psi$ for $\psi$ from a dense subset of $\HH_f$.
By \Cref{thm:Pauli}, the set of all $\psi=(H_0+i)\varphi$ with $\varphi \in C^{\infty}_0(\Omega) \cap \HHf$ is dense in $\HHf$, and for such $\psi$,
\begin{equation*}
      \| (\Hepsf+i)^{-1} \psi - R_0(i)\psi \| \leq  \Big\| \lambda_{\eps} \sum_{i<j}V_{\eps,ij}\varphi\Big\|
               \leq \lambda_{\eps} {N\choose 2}\|V_{\eps,12}\varphi\|,
\end{equation*}
where the (anti-)symmetry of $\ph$ was used in the second inequality. Let $\widetilde\varphi := \KK\varphi$, that is
$$
     \widetilde\varphi(r,R,x')=\varphi(R-r/2,R+r/2,x'),
$$
where $x' = (x_3,\ldots,x_N)$. Like $\ph$, $\tilde\ph$ is compactly supported and hence $\supp  \widetilde\varphi \subset \R^d \times B_{N-1}$ for some ball $B_{N-1}\subset \R^{d(N-1)}$. It follows that, for any $c>0$,
\begin{align}
    \eps^{4-d}  \|V_{\eps,12} \varphi \|^2 =& \int\limits_{B_{N-1}}  \d{R} \, \d{x'} \int\limits_{|r| \leq c} \left|V(r)\right|^2
            \left| \widetilde \varphi(\eps r,R,x') \right|^2\d{r} \nonumber \\
&+ \int\limits_{B_{N-1}}  \d{R} \,\d{x'} \int\limits_{|r| > c} \left|V(r)\right|^2\left| \widetilde \varphi(\eps r,R,x') \right|^2\d{r}. \label{intdecomp}
\end{align}
By assumption on $\ph$, $\tilde\ph(r,R,x') = 0$ for $r < \dist(\supp \varphi, \Gamma)$. This means that the first summand vanishes for $\eps c < \dist(\supp \varphi, \Gamma)$ and that 
$$
     \left| \widetilde \varphi(r,R,x') \right| \leq C(\widetilde \varphi,s) |r|^{s} 
$$
for each $s\geq 0$. It follows that 
$$
   \lambda_\eps  \|V_{\eps,12} \varphi \| \leq  \lambda_\eps \eps^{s+d/2-2} C(\widetilde \varphi,s) |B_{N-1}|^{1/2} \left( \int_{|r| > c} \left|V(r)\right|^2 |r|^{2s} \d{r}\right)^{1/2},
$$
where the integral can be made arbitrarily small by choosing $c$ large. By assumption $\limsup_{\eps \to 0}\lambda_\eps \eps^{s+d/2-2}<\infty$, hence it follows that $\lim_{\eps \to 0} \lambda_{\eps} \|V_{\eps,12}\varphi \| = 0$, and the proof is complete.
\end{proof}


\section{Examples and discussion}
\label{sec:four}

To put \Cref{thm:main} into a broader perspective and to demonstrate its dependence on the Pauli principle, we now view $H_\eps$ as the restriction
$$
        H_\eps = \tilde{H}_\eps\restricted\HH_f,
$$
where $\tilde{H}_\eps$ denotes the Schr\"odinger operator defined by expression \eqref{schrodinger} on the enlarged Hilbert space $L^2(\R^{Nd})$. We shall give choices for $\lambda_\eps$ and $V$, where $\tilde{H}_\eps$, in contrast to $H_\eps$, has a limit  $\tilde{H}$ describing non-trivial contact interactions or no limit at all.

In the cases $d=1$ and $d=2$ we choose, for simplicity, a two-body potential  
$V\in L^{\infty}(\R^d)$ with compact support and $\int V(r)\, dr=1$. Suppose further that 
\begin{alignat*}{2}
  \lambda_\eps &=g\eps >0 &&  \text{if}\ d=1,\\
  \lambda_\eps^{-1} &= \frac{|\log(\eps)|}{4\pi} + a\qquad && \text{if}\ d=2.
\end{alignat*}
Then $\lambdamax>0$ and $\lambda_0 = \limsup_{\eps\to 0} \lambda_\eps = 0$. So, the hypotheses of \Cref{thm:main} are satisfied and hence $H_\eps\to H_0$ in norm resolvent sense. On the other hand, by \cite{GHL2019,GH2021}, $\tilde{H}_\eps\to \tilde{H}$, where $\tilde{H}$ describes non-trivial contact interactions. That is, $\tilde{H}$ is a self-adjoint extension of $-\Delta\restricted C_0^{\infty}(\R^{Nd}\backslash\Gamma)$ distinct from $-\Delta$. See \eqref{def-Gamma} for the definition of $\Gamma$.

We now turn to the more interesting case of $N$ particles in $d=3$ dimensions. In the following $N$ is exhibited in the notation: we write $H_{N,\eps}$ for $H_{\eps}$ and $\tilde{H}_{N,\eps}$ for $\tilde{H}_{\eps}$. For the coupling constant and the two-body potential we choose
$\lambda_\eps = 2$ and
\begin{equation}
  V(r):= \begin{cases}
  \dfrac{2}{|r|}-1    \qquad\quad & \textup{if}\;\, |r| \leq 1 \\
  0  &\textup{if}\;\, |r| > 1.
\end{cases}\label{Vchoice}
\end{equation}
Then $0\leq V(r)\leq |r|^{-2}$ and hence $C_V=\sup V(r)|r|^2 \leq 1$. It follows that,
\begin{gather*}
   \lambda_0 = \limsup_{\eps\to 0} \lambda_\eps = 2\\
   \lambdamax \geq \frac{d^2}{NC_V} \geq \frac{9}{N}.
\end{gather*}
Thus, for $N\leq 4$ we have $\lambda_0<\lambdamax$ and hence, by \Cref{thm:main}, $H_{N,\eps}\to H_0$ in norm resolvent sense. On the other hand, concerning $\tilde{H}_{N,\eps}$ the following can be said:

\begin{prop}\label{thomas}
With the above notations, in the case $d=3$ we have
\begin{itemize}
\item[(a)] For $N=2$, $\tilde{H}_{2,\eps} \to \tilde{H}_{2}$ in norm resolvent sense, where $\tilde{H}_{2}$ is a non-trivial self-adjoint extension of $-\Delta\restricted C_0^{\infty}(\R^6\backslash\Gamma)$.
\item[(b)] For each $N\geq 3$ there exists a constant $C_N<0$ such that
$$
        \sigma(\tilde{H}_{N,\eps}) = [C_N\eps^{-2},\infty).
$$
\end{itemize}
\end{prop}

\noindent
\emph{Remark.} The divergence of the ground state energy established in Part (b) is known as Thomas effect \cite{Thomas1935}.

\begin{proof}
With respect to center of mass and relative coordinates $R=(x_1+x_2)/2$ and $r=x_2-x_1$, the Schr\"odinger operator for $N=2$ takes the form
\begin{align*}
    \tilde{H}_{2,\eps} &= -\Delta_R/2 \otimes 1 + 1\otimes h_\eps\\
     h_\eps &=  -2\Delta_r - \lambda_\eps V_\eps.
\end{align*}
By construction of $V$, $h=-\Delta -V\geq 0$ and $z=0$ is not an eigenvalue but a resonance energy. This means that the Birman-Schwinger operator $V^{1/2}(-\Delta)^{-1}V^{1/2}$ has the (simple) eigenvalue $1$, but the corresponding solution $\psi$ of $(-\Delta -V)\psi=0$ fails to be square integrable. Explicitly, in the present case, $\psi(x) = e^{-|x|}$ for $|x|\le 1$ and $\psi(x) = e^{-1}/|x|$ for $|x|>1$. These properties of $V$ imply that $h_\eps\to -2\Delta_0$ in norm resolvent sense, as $\eps\to 0$, where $\Delta_0$ denotes a self-adjoint extension of $\Delta\restricted C_0^{\infty}(\R^3\backslash\{0\})$ that is distinct from the free Laplacian \cite{SolvableModels}. Since $-\Delta_R/2\geq 0$, it follows that $\tilde{H}_{2,\eps} \to  -\Delta_R/2 \otimes 1 + 1\otimes  (-2\Delta_0)$ in norm resolvent sense, which proves assertion (a) (for details see \cite{Hofacker-Diss}).

In the case (b), we use that the Schr\"odinger operator $\tilde{H}_{N,\eps} $ is unitarily equivalent to $\eps^{-2}\tilde{H}_{N,\eps=1}$. For $N=3$ the presence of a zero-energy resonance in the two-body Hamiltonian leads to non-empty (in fact, infinite) discrete spectrum in the three-particle Hamiltonian with center of mass motion removed \cite{Efimov1970,Yafaev1974,OvchiSigal}.
This is the Efimov effect. It means, in particular, that $C_3:=\inf\sigma(\tilde{H}_{3,\eps=1})<0$. By the HVZ-theorem, $C_N:=\inf\sigma(\tilde{H}_{N,\eps=1})\leq C_3$ for all $N\geq 3$. 
\end{proof}


\section{Absence of contact interactions for $d\geq 2$}
\label{sec:appendix}

In space dimensions $d\geq 2$ zero-range interactions among equal-spin fermions are prohibited by the Pauli principle. This is true in the very strong form of Theorem \ref{thm:Pauli}, below. For a related result in the physics literature concerning two fermions in $d=2$, see \cite{BouSor}.

Let $\Gamma_{ij}:= \{x=(x_1,\ldots,x_N)\in \R^{Nd} \mid x_i=x_j\}$ and $\Omega_{ij} = \R^{Nd}\backslash \Gamma_{ij}$. Recall from Section~\ref{sec:three} that $\Gamma = \cup_{i<j}\Gamma_{ij}$ and 
$$
     \Omega = \R^{Nd}\backslash \Gamma = \bigcap_{i<j}\Omega_{ij}.
$$

\begin{thm}\label{thm:Pauli}
If $d\geq 2$, then $C_0^{\infty}(\Omega)\cap \HH_f$ is dense in $H^2( \R^{Nd}) \cap \HH_f$ with respect to the norm of $H^2$. This means that 
$$
        H^2_0( \Omega) \cap \HH_f =  H^2( \R^{Nd}) \cap \HH_f,
$$
and it implies that the Laplacian $\Delta$ in $\HH_f$ is essentially self-adjoint on $C_0^{\infty}(\Omega)\cap\HH_f$.
\end{thm}

\noindent
\emph{Remarks.}  
\begin{enumerate}
\item The main point of \Cref{thm:Pauli} is that elements of $C_0^{\infty}(\Omega)$ vanish in an entire neighborhood of the collision set $\Gamma$. The elements of $C_0^{\infty}(\R^{Nd})\cap \HH_f$ vanish on $\Gamma$ too. But
the weaker statement, that $C_0^{\infty}(\R^{Nd})\cap \HH_f$ is dense in $H^2( \R^{Nd}) \cap \HH_f$, is true for all $d\geq 1$ and it easily follows from the fact that $C_0^{\infty}(\R^{Nd})$ is dense in $H^2( \R^{Nd})$.

\item For $d=1$ the assertion of the theorem is false. To see this, consider a sequence $(\psi_n)$ in $C_0^{\infty}(\Omega)$ with $\psi_n\to \psi$ in the norm of $H^2$. Then $\nabla\psi_n\to \nabla\psi$ in the norm of $H^1$. Since the trace operators  $T_{ij}: H^1(\R^N) \to L^2(\Gamma_{ij})$ are continuous, and since, clearly, $\nabla\psi_n = 0$ on all hyperplanes $\Gamma_{ij}$, it follows that
\begin{equation}\label{zero-trace}
      \nabla\psi = 0\quad \text{on all}\ \Gamma_{ij},
\end{equation}
or, more precisely, $T_{ij}\nabla\psi = 0$ in $L^2(\Gamma_{ij})$.
We now give an example of an antisymmetric wave function $\psi\in H^2(\R^N)$ without property \eqref{zero-trace}, which proves that $C_0^{\infty}(\Omega)\cap \HH_f$ is not dense in $H^2(\R^{N}) \cap \HH_f$.

Let $|x|^2:=\sum_{i=1}^N x_i^2$ and 
$$
      \psi(x_1,\ldots, x_N) :=  e^{-|x|^2} \prod_{i<j} (x_j-x_i).
$$
Apart from the Gaussian, this is a Vandermonde determinant. This shows that $\psi$ is antisymmetric. On the hyperplane $\Gamma_{12}$ we have 
$$
    \frac{\partial\psi}{\partial x_1}\Big|_{x_1=x_2} = - e^{-|x|^2}\prod_{j=3}^N (x_j-x_1)\prod_{2\leq i<j} (x_j-x_i),
$$
which shows that $\nabla\psi$ does not vanish on $\Gamma_{12}$.
\end{enumerate}

The proof of \Cref{thm:Pauli} is based on the following lemmas, \Cref{lm1} being the heart of it.


\begin{lemma}\label{lm1}
If $d\geq 2$, then there exists a sequence $u_n\in C_0^{\infty}(\R^d,[0,1])$ with $u_n(x)=1$ if $|x|\leq 1/n$, and, in the limit $n\to\infty$, $\diam(\supp u_n)\to 0$ as well as
\begin{gather}
 \int |\nabla u_{n}(x)|^2\, dx \to 0\\
 \int |x|^2|\Delta u_{n}(x)|^2\, dx \to 0.
\end{gather}
\end{lemma}

\begin{proof}
In the case $d\geq 3$ we may choose any function $u\in C_0^{\infty}(\R^d,[0,1])$ with $u(x)=1$ for $|x|\leq 1$ and define $u_n(x)=u(nx)$. Then, with the substitution $y=nx$, in the limit $n\to\infty$,
\begin{align*}
      \int |\nabla u_{n}(x)|^2\, dx &=  \int |\nabla u(y)|^2\, dy\cdot n^{2-d} \to 0,\\ 
      \int |x|^2|\Delta u_{n}(x)|^2\, dx &=  \int |y|^2|\Delta u(y)|^2\, dy\cdot n^{2-d} \to 0.  
\end{align*}
In the case $d=2$ we define $u_n(0):=1$ and for $|x|>0$ we set
$$
     u_n(x) := g\left(\frac{\log(n|x|)}{\log\log n}\right),
$$
where $g\in C^{\infty}(\R,[0,1])$ with 
$$
        g(s) := \begin{cases}1 & s\leq 0\\ 0 & s\geq 1. \end{cases}
$$
It follows that $u_n(x)=1$ for $|x|\leq 1/n$, $u_n(x)=0$ for $|x|\geq (\log n)/n$ and hence that  $u_n\in C_0^{\infty}(\R^d,[0,1])$. Moreover,
\begin{align*}
      \frac{1}{2\pi} \int |\nabla u_{n}(x)|^2\, dx &= \int_{1/n}^{(\log n)/n}  g'\left(\frac{\log(nr)}{\log\log n}\right)^2 \frac{dr}{r}\cdot  \frac{1}{(\log\log n)^2}\\
     &=\int_0^1 g'(s)^2\, ds\cdot  \frac{1}{\log\log n}.
\end{align*}
On the other hand, using that on radially symmetric functions
$$
   r^2\Delta = \left(r\frac{\partial}{\partial r}\right)^2,
$$
we find,
\begin{align*}
      \frac{1}{2\pi}  \int |x|^2|\Delta u_{n}(x)|^2\, dx  &= \int_{1/n}^{(\log n)/n}  \left| \left(r\frac{\partial}{\partial r}\right)^2g\left(\frac{\log(nr)}{\log\log n}\right)\right|^2 \frac{dr}{r}\\ 
       &= \int_{1/n}^{(\log n)/n}  g''\left(\frac{\log(nr)}{\log\log n}\right)^2 \frac{dr}{r} \cdot  \frac{1}{(\log\log n)^4}\\
       &=\int_0^1 g''(s)^2\, ds\cdot  \frac{1}{(\log\log n)^3}.
\end{align*}
This concludes the proof.
\end{proof}

\begin{lemma}\label{lm:iteration-step}
Suppose that $d\geq 2$ and let $\psi\in C_0^{\infty}(\R^{Nd})$ with $\psi = 0$ on $\Gamma$. Then,  for each pair $i,j\in \{1,\ldots,N\}$, $i\neq j$ and  for each $\eps>0$, there exists $\psi_\eps\in C_0^{\infty}(\Omega_{ij})$ with $ \supp\psi_\eps\subset\supp\psi$, $\psi_\eps = 0$ on $\Gamma$, and
$$
       \|(-\Delta+1)(\psi-\psi_\eps)\| < \eps.
$$
\begin{proof}
We may assume that $(i,j) = (1,2)$ and we introduce the relative and center of mass coordinates 
$$
     r:= x_2-x_1,\qquad R:=\frac{1}{2}(x_1+x_2).
$$
Then 
\begin{equation}\label{r-Lap}
      \Delta  = 2 \Delta_r + \frac{1}{2}\Delta_R + \Delta_{x'},
\end{equation}
where $x':=(x_3,\ldots,x_N)$. Let $\psi\in C_0^{\infty}(\R^{Nd})$ with $\psi = 0$ on $\Gamma$ and let 
$$
     \psi_n(x_1,\ldots,x_N) =\psi(x_1,\ldots,x_N)\cdot (1-u_n(x_2-x_1))
$$
where $u_n$ is given by Lemma~\ref{lm1}. In the following $u_n$ also denotes the function $(x_1,\ldots,x_N) \mapsto u_n(x_2-x_1)$. 
Then $\psi-\psi_n = \psi u_n$ and hence
$$
    \|(-\Delta+1)(\psi-\psi_n)\|  \leq \|\psi u_n\| + \|\Delta (\psi u_n)\|.  
$$
Clearly $ \|\psi u_n\| \to 0$ because $|\psi u_n| \leq |\psi|$ and because $\psi u_n\to 0$ pointwise as $n\to\infty$. On the other hand, using \eqref{r-Lap} and the fact that $u_n$ depends on $r$ only,
\begin{align*}
     \Delta(\psi u_n)& = 2 \Delta_r(\psi u_n) + \frac{1}{2}(\Delta_R\psi)u_n + (\Delta_{x'}\psi)u_n
\end{align*}
where the first term equals
$$
       2\Delta_r(\psi u_n) = 2(\Delta_r\psi)u_n + 4(\nabla_r\psi)(\nabla_r u_n) + 2\psi \Delta_ru_n.
$$
By the pointwise convergence $u_n\to 0$, as explained above, $(\Delta_r\psi)u_n, (\Delta_R\psi)u_n$ and $(\Delta_{x'}\psi)u_n$ have vanishing $L^2$-norm in the limit $n\to\infty$. It remains to show that 
\begin{align*}
     \|\nabla_r\psi\cdot \nabla_r u_n\| &\to 0\qquad (n\to\infty)\\
      \|\psi\cdot \Delta_r u_n\| &\to 0\qquad (n\to\infty).
\end{align*}
From Lemma~\ref{lm1} we know that 
\begin{align*}
      \|\nabla_r\psi\cdot \nabla_r u_n\|^2 &\leq \int |\nabla_r\psi(r,R,x')|^2 |\nabla u_n(r)|^2\, drdRdx'\\ 
   &\leq  \sup_{r\in\R^d} \int |\nabla_r\psi(r,R,x')|^2\, dR dx' \cdot \|\nabla u_n\|^2
   \quad \to 0\qquad (n\to\infty). 
\end{align*}
For $\psi \Delta u_n$ we use that $\psi(0,R,x') = 0$ and hence that 
$$
    \psi(r,R,x') = \int_0^1(\nabla_r\psi)(tr,R,x')\cdot r\, dt.
$$
It follows that 
\begin{eqnarray*}
  \lefteqn{\int |\psi\Delta u_n|^2\, drdRdx'}\\
  && \leq \int dr dR dx' \left(\int_0^1 |\nabla_r\psi(tr,R,x')|^2\,dt\right)|r|^2 |\Delta u_n(r)|^2\\
  && \leq C \int |r|^2  |\Delta u_n(r)|^2\, dr \to 0,\qquad (n\to\infty)
\end{eqnarray*}
by \Cref{lm1}, because
$$
    C:= \sup_{r\in\R^d} \int dRdx' \int_0^1 |\nabla_r\psi(tr,R,x')|^2\,dt <\infty. \qedhere
$$
\end{proof}
\end{lemma}

\begin{proof}[Proof of \Cref{thm:Pauli}]
For given $\psi \in H^2(\R^{Nd}) \cap \HH_f$ and $\eps>0$ it suffices to find $\phi_\eps \in C_0^{\infty}(\Omega) $ with $\|\psi -\phi_\eps\|_{H^2}<\eps$. Let $P_f: L^2(\R^{Nd})\to L^2(\R^{Nd})$ denote the orthogonal projection onto $\HH_f$.
Then $\psi_{\eps}:= P_f \phi_\eps$ belongs to $C_0^{\infty}(\Omega)\cap \HH_f$ and 
\begin{align*}
    \|\psi - \psi_\eps\|_{H^2}  &= \|P_f(\psi - \phi_\eps)\|_{H^2} \leq     \|\psi - \phi_\eps\|_{H^2}  <\eps
\end{align*}
because $P_f$ is an orthogonal projection in $H^2$ (if suitably normed). To find $\phi_\eps$, we may assume that $\psi\in C_0^{\infty}(\R^{Nd})\cap \HH_f$, which is dense in  $H^2(\R^{Nd}) \cap \HH_f$, and we use \Cref{lm:iteration-step} repeatedly. That is, we  use $\{\sigma_k\mid k=0,\ldots,n\}$ to denote the set of $n:={N\choose 2}$ pairs $(i,j)$, we define $\Omega_0:=\R^{Nd}$ and 
$$
      \Omega_k:= \cap_{j=1}^k \Omega_{\sigma_j},\qquad k=1\ldots n.
$$
Then we construct smooth functions $(\gamma_k)_{k=0}^n$ recursively with $\gamma_0:= \psi$, $\supp(\gamma_k)\subset \supp(\gamma_{k-1}) \cap \Omega_k$, $\gamma_k=0$ on $\Gamma$, and $\|(-\Delta+1)(\gamma_k-\gamma_{k-1})\| <\eps/n$. This is achieved with the help of 
\Cref{lm:iteration-step}. The function $\phi_\eps:= \gamma_n$ has the desired properties.
\end{proof}

\noindent
\textbf{Acknowledgement.} We thank Semjon Wugalter for pointing out reference \cite{KlausSimon1980}.






\end{document}